\documentstyle[11pt,newpasp,twoside,epsf]{article}
\def\edcomment#1{\iffalse\marginpar{\raggedright\sl#1\/}\else\relax\fi}
\reversemarginpar

\begin{document}
\title{Evolution of the X-ray Properties of Clusters of Galaxies}
 \author{J. Patrick Henry}
\affil{Institute for Astronomy, University of Hawaii, 2680 Woodlawn Drive,
Honolulu, HI 96822, USA}

\begin{abstract}
The amount and nature of the evolution of the X-ray properties of
clusters of galaxies provides information on the formation of
structure in the universe and on the properties of the universe
itself.  The cluster luminosity - temperature relation does not evolve
strongly, suggesting that the hot X-ray gas had a more complicated
thermodynamic history than simply collapsing into the cluster
potential well. Cluster X-ray luminosities do evolve. The dependence
of this evolution on redshift and luminosity is characterized using
two large high redshift samples. Cluster X-ray temperatures also
evolve. This evolution constrains the dark matter and dark energy
content of the universe as well as other parameters of cosmological
interest.
\end{abstract}

\section{Introduction}

The evolution of cosmic structure is strongly dependent on the
cosmology of the universe. Jenkins et al. (1998) is one of the many
papers describing this well known result. Clusters of galaxies, as the
most massive bound objects known, are the ultimate manifestations of
cosmic structure building. The evolution of clusters is simple, being
driven by the gravity of the underlying mass field of the universe and
of a collisionless collapse of cluster dark matter. It should be
possible to calculate this evolution reliably compared to that of
other objects visible at cosmological distances such as galaxies or
AGN. Clusters are luminous X-ray sources. The X-ray emission mechanism
is optically thin thermal radiation from a medium nearly in
collisional equilibrium, about the simpliest situation imaginable.
Thus observations of the X-ray evolution of clusters provide a robust
measure of the evolution of cosmic structure and thereby constrain the
cosmology of the universe.

The cosmological model is described by a set of cosmological
parameters. The present value of the Hubble parameter is $\mathrm{H_0
\equiv 100\:h\:km\:s^{-1}\:Mpc^{-1}}$. When needed h = 0.5 will be
used, but almost nothing about evolution depends on the precise value
of h since data at two epochs are always compared. The present matter
and dark energy densities in terms of the critical density are
$\Omega_{m0}$ and $ \Omega_{\Lambda0}$ respectively. The amount of
structure in the universe is described by $\sigma_8$, the present rms
matter fluctuations in spheres of $\mathrm{8\:h^{-1}\:Mpc}$. This
parameter is a complicated way to describe the present normalization
of the spatial power spectrum of matter density fluctuations, P(k), on
a scale of k $\sim\:0.2 \mathrm{\:h\: Mpc^{-1}}$: $\mathrm{\sigma_8
\approx [P(0.172\:h\:Mpc^{-1})/3879\:h^{-3}\:Mpc^3]^{1/2}}$ (Peacock, 1999
equations 16.13 and 16.132). The dark energy equation of state is P =
w $\rho$ c$^2$. If w = -1, then the dark energy is the cosmological
constant, if $\mathrm{-1 < w < 0}$ it is termed Quintessence. Recall
that w = 0 is cold dark matter and w = 1/3 is radiation. Cluster
temperature evolution provides constraints on all of the above
cosmological parameters except h.

When needed, two specific cases will be considered. The X-ray astronomer's
universe where $\Omega_{m0} = 1.0, \Omega_{\Lambda0} = 0.0$. This combination
was known to be correct ten years ago. The other case may be called the
bandwagon universe where $\Omega_{m0} = 0.3, \Omega_{\Lambda0} = 0.7$.
This combination is known to be correct today. Both universes are
spatially flat, $\Omega_{m0} + \Omega_{\Lambda0} = 1$.

\section{Cluster Luminosity - Temperature Relation Evolution}

The relation between cluster X-ray luminosities and temperatures (L -
T relation) is the oldest and best studied among cluster X-ray
observables. Early studies include Mitchell, Ives \& Culhane (1977),
Mushotzky et al. (1978) and Henry \& Tucker (1979). Recent
comprehensive work includes David et al. (1993), White, Jones \&
Forman (1997), Markevitch (1998) and Arnaud \& Evrard (1999). Figure 1
shows an example of this relation using a sample of 53 $z < 0.3$ and
32 $0.3\leq z \leq 0.6$ clusters all with ASCA temperatures (Novicki,
Sornig, \& Henry, 2002).

\begin{figure}
\plottwo{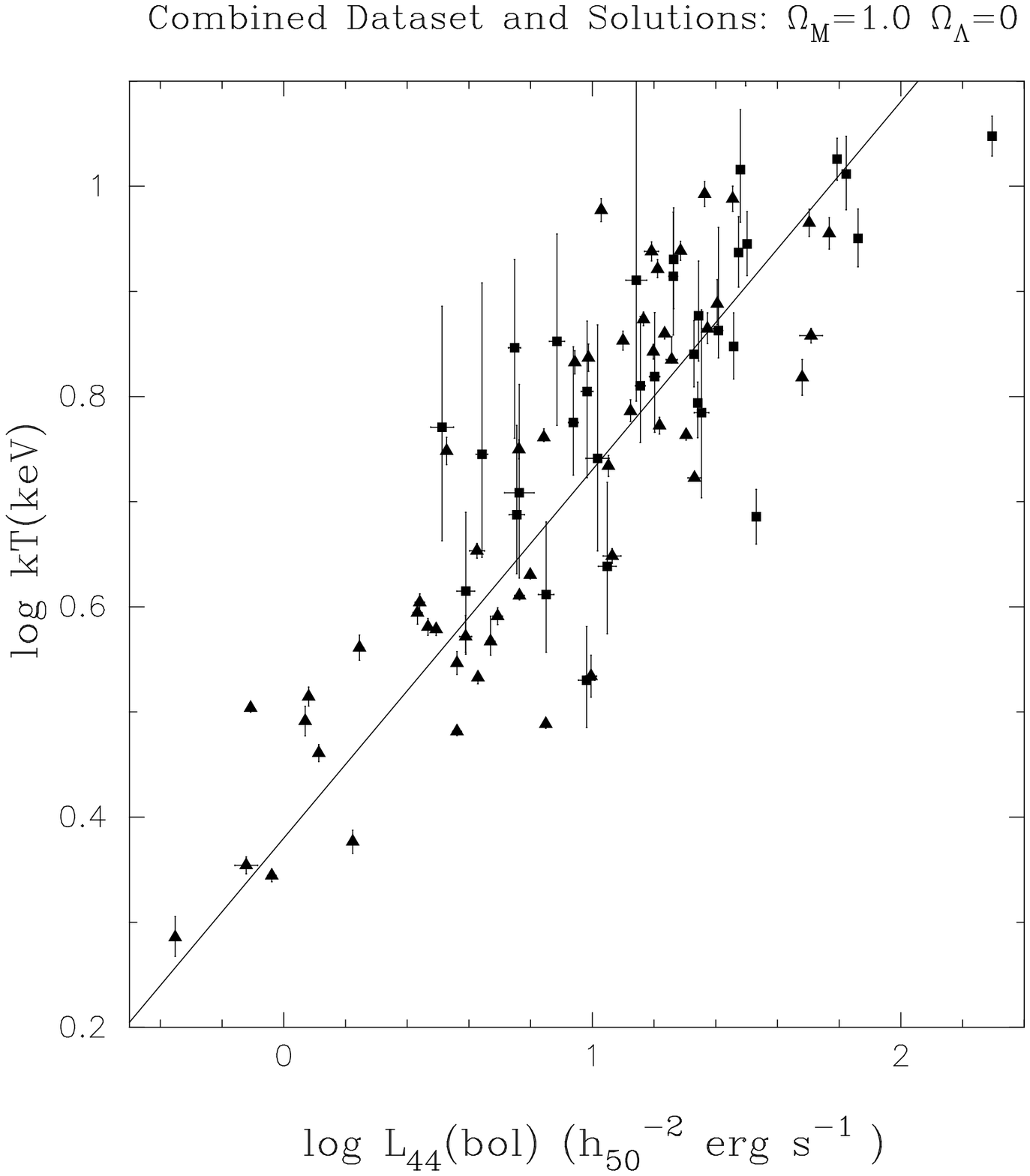}{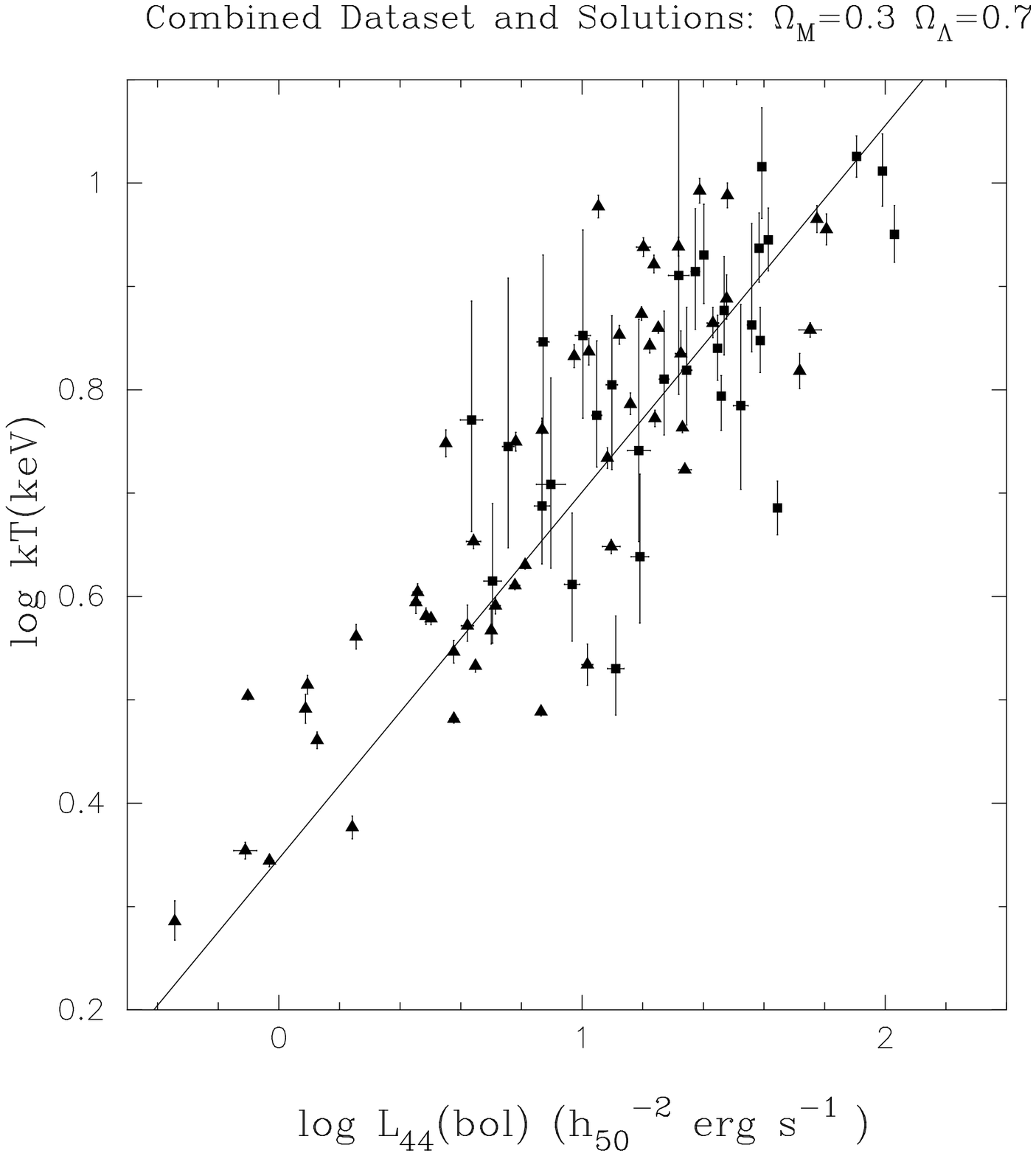}
\caption{The cluster luminosity - temperature relation. Triangles are the
low redshift sample and squares are the high redshift sample. The line
is the best fit to the entire sample of $L(bol) = C[kT]^{\alpha}(1 +
z)^{A}$. a. (left) For a cosmology with $\Omega_{m0} = 1.0,
\Omega_{\Lambda0} = 0.0$. b. (right) For a cosmology with $\Omega_{m0}
= 0.3, \Omega_{\Lambda0} = 0.7$.}
\end{figure}

Although there is clearly a relation in Figure 1, there is also
substantial non-statistical scatter. The scatter is approximately
Gaussian distributed in the logarithm of the luminosity at constant
temperature.  The dispersion in the logarithm of the bolometric
luminosity, $\mathrm{\sigma logL(bol)}$, is about 0.2. The scatter
is reduced if the effects of the cluster cooling centers are removed,
either by modeling them (Allen \& Fabian, 1998), excising them
(Markevitch, 1998), or only using clusters that do not have cooling
centers (Arnaud \& Evrard, 1999).

Most observers adopt the simple functional form $L(bol) =
C[kT]^{\alpha}(1 + Z)^{A}$, where $L(bol)$ is the bolometric
luminosity, to describe their results. The observations indicate that
$\alpha \sim 3$. 
However, simple scaling relations predict $\alpha = 2$ and $A = 1.5$.
This $\alpha$ discrepancy is usually resolved by positing that the
cluster gas has experienced some form of preheating prior to its
heating through gravitational collapse (Evrard \& Henry, 1991; Kaiser,
1991; Bialek, Evrard, \& Mohr 2001; Tozzi \& Norman, 2001; Bower et
al., 2001). Preheating raises the temperatures of low mass clusters
more than those of high mass, thereby breaking the scaling
relations. These preheating models predict little or no evolution of
the L - T relation, at least to redshifts $\sim 1$.  For example, the
model in Evrard \& Henry (1991) has $\alpha = 2.5$ and $A = 0$. An
alternative to this idea invokes cooling, which is more efficient at
lower temperatures, and a small amount of post collapse heating to
break the scaling (Voit \& Bryan, 2001). Little evolution is expected
in this model as well. Thus the values of $\alpha$ and $A$ provide
clues to the thermodynamic history of the cluster gas.

\begin{figure}
\plottwo{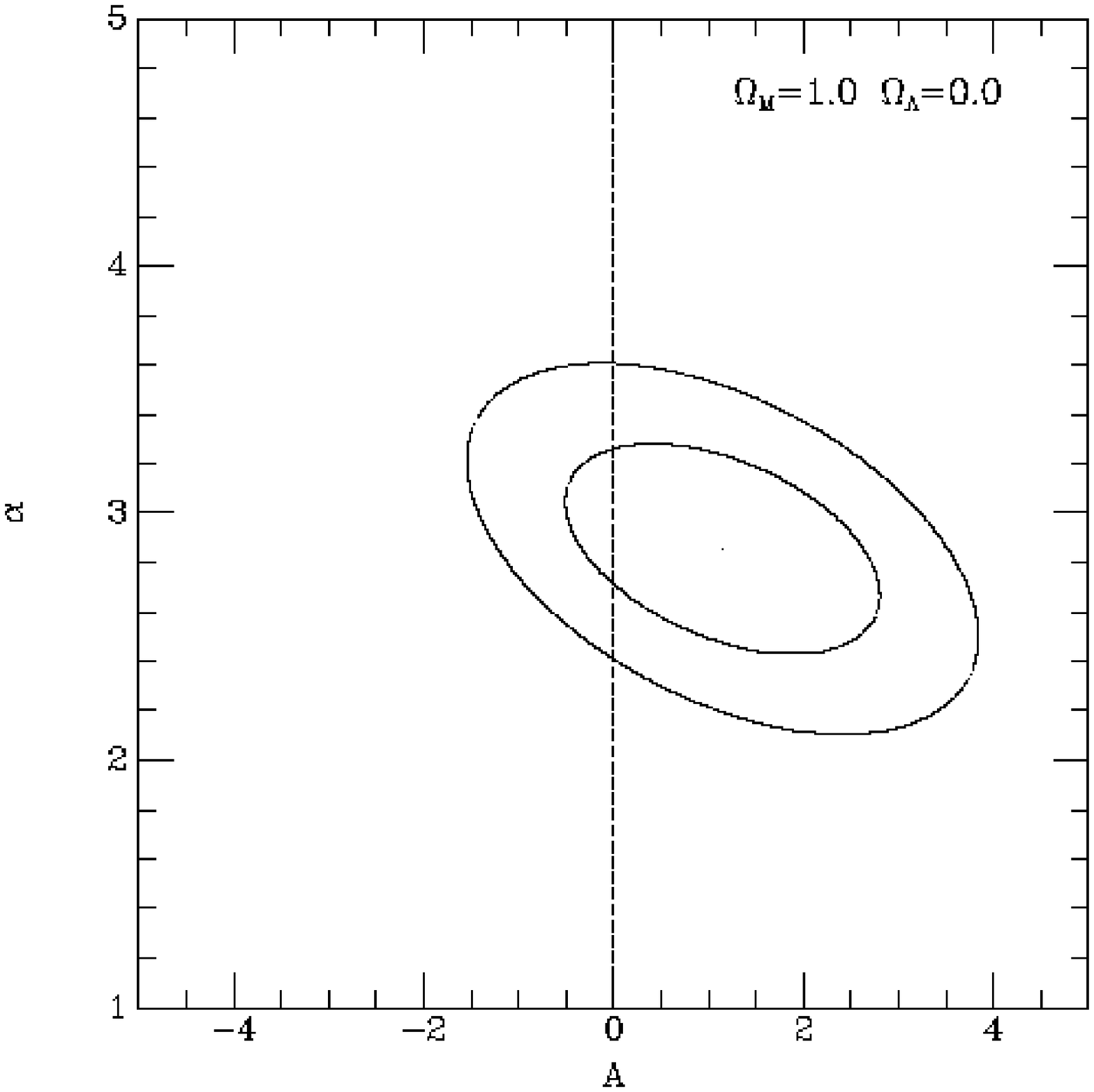}{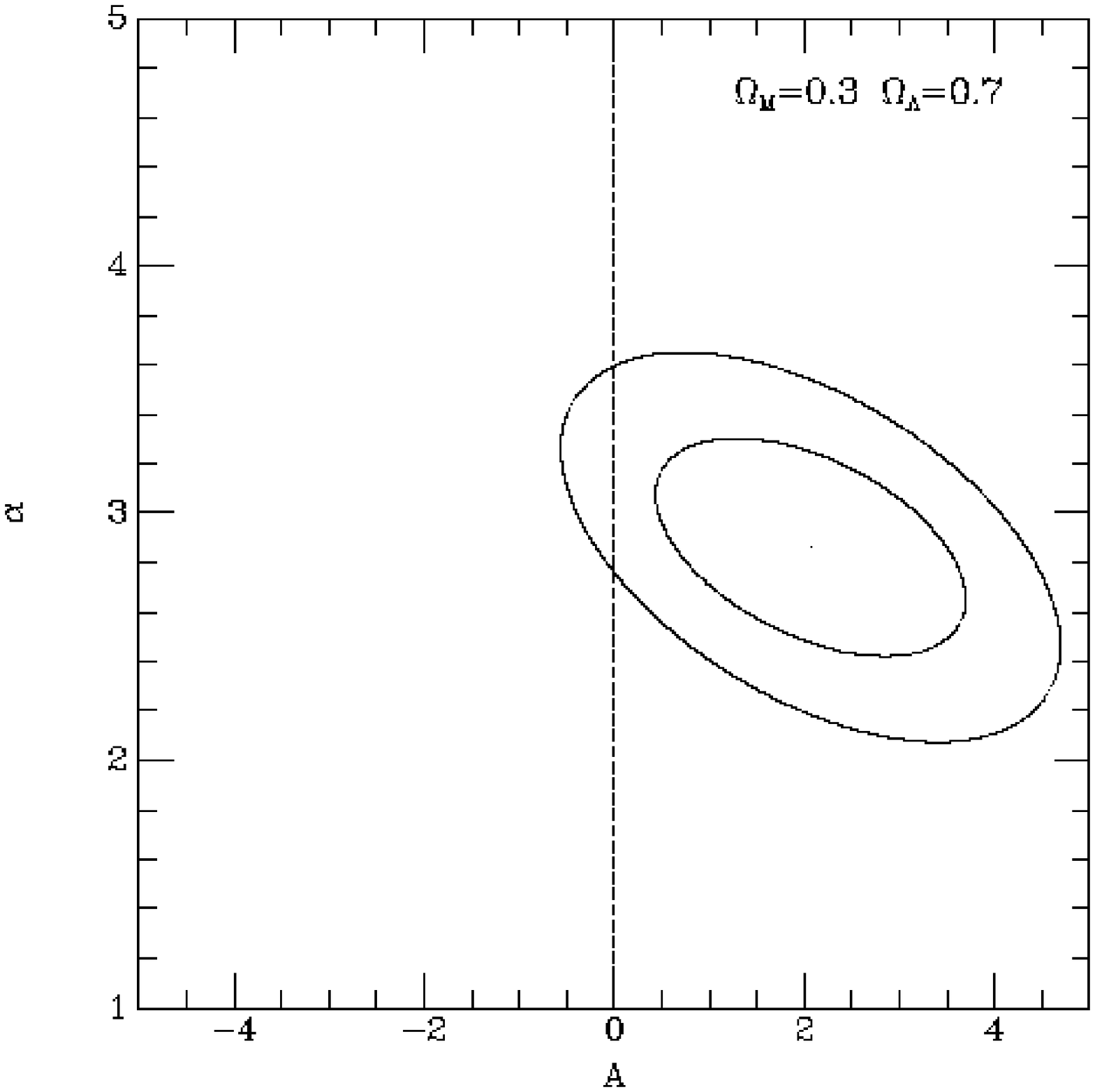}
\caption{The 1 and 2 $\sigma$ contours for a fit of the data in Figure 1
to $L(bol) = C[kT]^{\alpha}(1 +z)^{A}$. a. (left) For a cosmology with 
$\Omega_{m0} = 1.0,\Omega_{\Lambda0} = 0.0$. b. (right) For a cosmology 
with $\Omega_{m0}= 0.3, \Omega_{\Lambda0} = 0.7$.}
\end{figure}

It has recently become possible to constrain the evolution of the L -
T relation, thereby testing these scenarios. Figure 2 shows
the constraints derived by Novicki et al. (2002). The evolution is
mild; A is consistent with 0 at the $\sim 1 \sigma$ level. Similar
results have been found by Henry, Jiao, \& Gioia (1994), Mushotzky \&
Scharf (1997), Sadat et al. (1998), Donahue et al. (1999), Reichart,
Castander \& Nichol (1999), Fairley et al. (2000), and Arnaud,
Aghanian \& Neumann (2002). Little or no evolution may then imply that
the thermodynamic history of cluster gas is more complicated than
simply falling into the cluster potential well, at least for low mass
clusters.  There was possibly some additional heating prior to
collapse and/or cooling after collapse.

\section{Cluster X-ray Luminosity Evolution}

The initial claim for cluster X-ray luminosity evolution was made over
a decade ago by Gioia et al. (1990) and Henry et al. (1992) using the
Einstein Extended Medium Sensitivity Survey (EMSS). The reported
evolution was in the sense that there are fewer high luminosity
clusters in the past.  This conclusion has not enjoyed universal
acceptance. However, there are now many new samples of clusters, both
at high and low redshift, with which it may be tested.

\subsection{Cluster X-ray Luminosity Samples}

Clusters are not standard luminosity candles. Therefore, searching for
luminosity evolution requires measuring the X-ray luminosity function
at two epochs and that requires constructing statistically complete
samples at the two epochs. Fortunately, measuring the X-ray luminosity
is relatively easy. It only requires $\sim 25$ photons for a $5
\sigma$ luminosity measurement, so large samples are possible with
relatively modest observing time.

{\small
\begin{table}
\caption{Low Redshift Cluster Luminosity Samples}
\begin{tabular}{lrllrl}
\tableline
Name&Num&Sky Cut     &F(0.1,2.4)(cgs)&$\Omega$(deg$^2$)   &Reference\\
\tableline
BCS  &201&$\delta > 0.0$&$>4.4\times10^{-12}$&13,578&Ebeling et al. (1998)\\
     &   &$\mid b \mid > 20$& & & \\
REFLEX&441&$\delta < 2.5$&$>3.0\times10^{-12}$&13,924&B\"ohringer et al. (2001)\\
     &   &$\mid b \mid > 20$& & & \\
\tableline
\tableline
\end{tabular}
\end{table}
}

Table 1 gives some properties of low redshift (here defined as $z <
0.3$) statistically complete cluster luminosity samples. Not included
in Table 1 are the RASS1 Bright Sample (De Grandi et al., 1999),
which is a subsample of REFLEX and the eBCS sample (Ebeling et al.,
2000), which adds another 100 clusters at $\mathrm F(0.1,2.4) >
2.8\times10^{-12}$ $\mathrm erg\: cm ^{-2}\: s^{-1}$ to the BCS but is
75\% complete. The REFLEX sample was constructed from the second
(i.e. fully merged) processing of the ROSAT All-Sky Survey (RASS2).

{\small
\begin{table}
\caption{High Redshift Cluster Luminosity Samples}
\begin{tabular}{lrllrl}
\tableline
Name&Num&Sky Cut     &F(0.5,2.0)(cgs)&$\Omega$(deg$^2$)   &Reference\\
\tableline
EMSS  &22&$\delta > -40$&$>4.6\times10^{-14}$&778&Gioia \& Luppino (1994)\\
     &   &$\mid b \mid > 20$& & & \\
MACS&119&$\delta < 80$&$>6.2\times10^{-13}$&22,735&Ebeling et al. (2001)\\
     &  &$\delta > -40$& & & \\
     &  &$\mid b \mid > 20$& & & \\
NEP  &19&$\alpha = 18$&$>2.2\times10^{-14}$&81&Henry et al. (2001)\\
     &   &$\delta = 66.5$& & & \\
160 deg$^2$&73&$\mid b \mid > 20$&$>1.3\times10^{-14}$&158&Vikhlinin et al. (1998)\\
RDCS3&50&$\mid b \mid > 20$&$>3.0\times10^{-14}$&47&Rosati (2001)\\
BSHARC&12&$\mid b \mid > 20$&$>1.6\times10^{-13}$&179&Romer et al. (2000)\\
WARPS II&78&$\mid b \mid > 20$&$>6.0\times10^{-14}$&73&Jones et al. (2001)\\
\tableline
\tableline
\end{tabular}
\end{table}
}

Table 2 gives some properties of the high redshift cluster luminosity
samples. Note that most of these samples also have low redshift
objects, but only the numbers of $z > 0.3$ objects are listed in the
table. The EMSS sample is from Einstein Observatory observations, the
other samples are from ROSAT. The MACS and NEP samples come from the
RASS2, i.e. they are contiguous, all other samples are from disjoint
targeted observations with a region around the target excised.

Typically, low redshift cluster samples contain several hundred X-ray
bright objects coming from one-third of the sky. High redshift samples
usually contain several tens of faint objects coming from less than
one-percent of the sky.

\subsection{Test For Cluster X-ray Luminosity Evolution In Seven High
Redshift Samples}

\begin{figure}
\plotfiddle{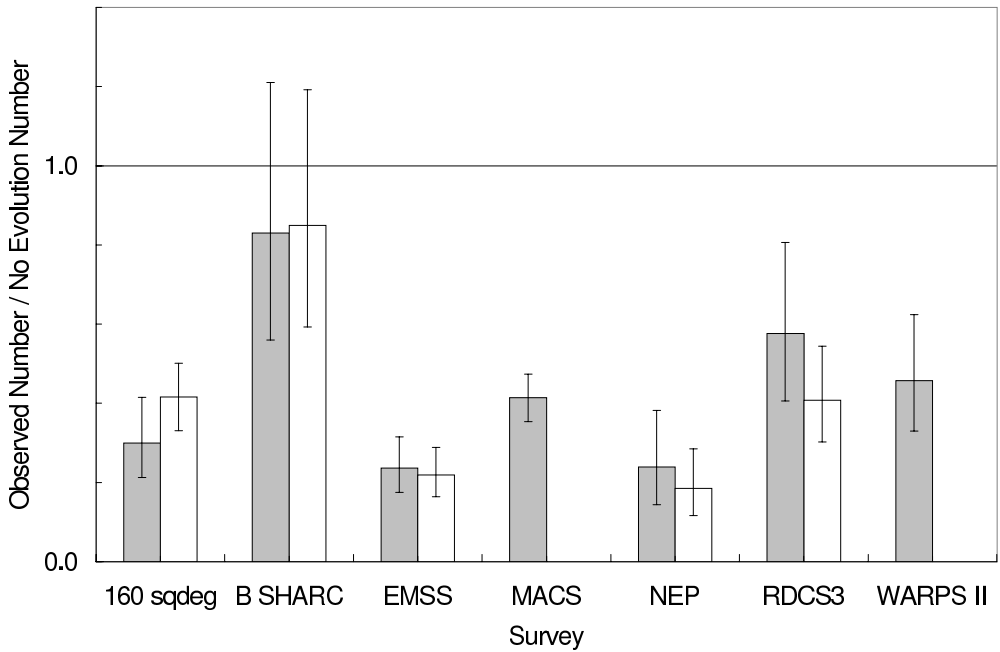}{3.3in}{-90}{57}{57}{-240}{290}
\caption{Ratio of the observed to no-evolution expectation numbers of
clusters for seven high redshift surveys. The grey bar assumes 
$\Omega_{m0} = 1.0$, $\Omega_{\Lambda0} = 0.0$; the
clear bar $\Omega_{m0} = 0.3$, $\Omega_{\Lambda0} = 0.7$. The ratio
can not yet be made for MACS and WARPS II for the latter case because
the individual cluster luminosities and redshifts are not yet published.}
\end{figure}

Since the evolution appears to be a lack of objects at high
luminosities, simply comparing the binned low and high redshift
differential luminosity functions will not use the full statistical
power of the sample. In the limit that the evolution is a downward
step function above a certain luminosity, such a comparison will show
no difference between the two epochs (unless upper limits are
plotted).

A more sensitive test is to compare the observed number of 
high-redshift, high-luminosity clusters to the expected number if the low
redshift luminosity function did not evolve. This test accounts for
clusters that could have been found but were not. 

Specifically, the test used here compares the number of observed
clusters with luminosities in the 0.5 - 2.0 keV band $\geq 2 \times
10^{44}$ $\mathrm {erg\: s^{-1}}$ (except MACS, whose selection
function limits the lowest luminosity to $2.9 \times 10^{44}$$\mathrm
{erg\: s^{-1}}$) and with $0.3 \leq z\leq 1.0$ to the expected number
assuming the REFLEX low redshift luminosity function did not
evolve. Each of the seven high redshift samples probes a different
region of the L - z plane due to its different flux limit and solid
angle. The average redshift and luminosity of the clusters after the
above cuts for all seven samples is 0.495 and $4.7 \times 10^{44}$
$\mathrm {erg\: s^{-1}}$ respectively ($\Omega_{m0} = 1.0$,
$\Omega_{\Lambda0} = 0.0$). Figure 3 summarizes the results.  All
seven samples show evolution, most at a statistically significant
level. The average ratio of the observed to no-evolution numbers is
$0.35 \pm 0.04$ for $\Omega_{m0} = 1.0$, $\Omega_{\Lambda0} = 0.0$,
and $0.29 \pm 0.04$ for $\Omega_{m0} = 0.3$, $\Omega_{\Lambda0} =
0.7$.

\begin{figure}
\plotfiddle{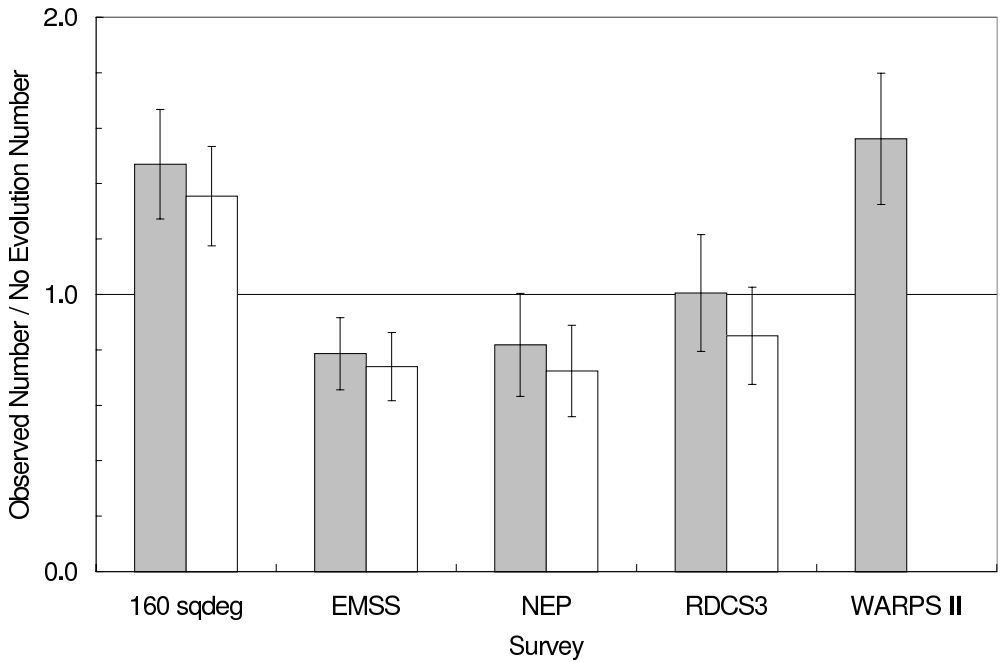}{3.2in}{-90}{57}{57}{-240}{290}
\caption{Ratio of the observed to no-evolution expectation numbers of
clusters for the low redshift components of five high-redshift
surveys. The sample definition is L(0.5,2.0) $\geq10^{43}$ $\mathrm
{erg\: s^{-1}}$ and $0.15 \leq z\leq 0.30$. The grey bar assumes
$\Omega_{m0} = 1.0$, $\Omega_{\Lambda0} = 0.0$; the clear bar
$\Omega_{m0} = 0.3$, $\Omega_{\Lambda0} = 0.7$.}
\end{figure}

Figure 3 also shows that one systematic effect, the assumed cosmology,
is not important. Another systematic effect is the choice of the
low-redshift luminosity function. Using the BCS instead of REFLEX
yields a ratio of $0.38 \pm 0.04$ for $\Omega_{m0} = 1.0$,
$\Omega_{\Lambda0}= 0.0$ (the BCS investigators only gave results for
this cosmology). Thus the choice of the low-redshift comparison
luminosity function is also not important. The final systematic
investigated was whether the test yields a null result, a ratio of 1,
at low redshift. Figure 4 shows that it does. The average ratio of the
observed to no-evolution numbers is $1.03 \pm 0.08$ for $\Omega_{m0} =
1.0$, $\Omega_{\Lambda0} = 0.0$, and $0.87 \pm 0.08$ for $\Omega_{m0}
= 0.3$, $\Omega_{\Lambda0} = 0.7$. This test is limited to five
samples because the selection function for Bright SHARC has not been
published at low redshift and MACS does not attempt to be complete at
low redshift.

These results show that there is no longer a question of whether
cluster X-ray luminosities evolve, they do. Seven nearly independent
surveys show some evolution with a combined significance $> 15\sigma$.
Systematic effects do not appear to be important. The amount of
evolution is about a factor of three at $L(0.5,2.0) \sim 4.7 \times
10^{44}$ $\mathrm {erg\: s^{-1}}$ over the redshift range $\sim0.1$ to
$\sim0.5$. It is now time to better characterize the evolution.

\subsection{Characterizing Cluster Luminosity Evolution Using Two Large
High Redshift Samples}

\begin{figure}
\plotfiddle{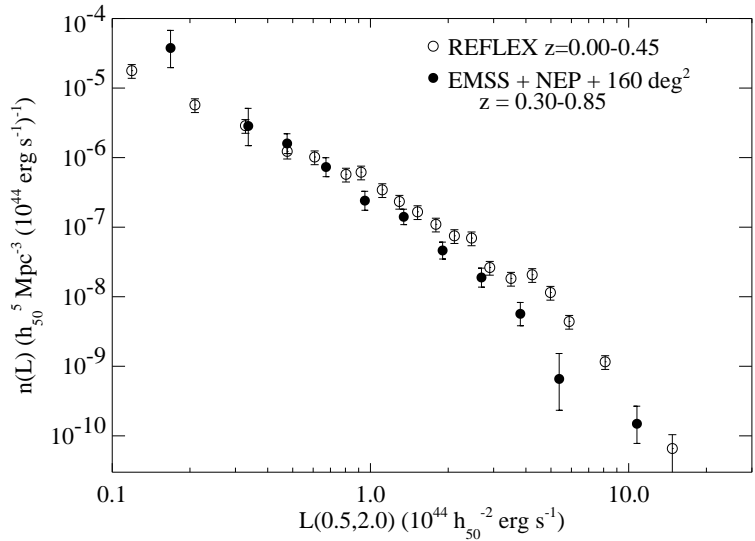}{2.5in}{0}{60}{60}{-190}{-230}
\caption{X-ray luminosity functions from the low-redshift REFLEX 
(from B\"ohringer et al., 2002) and combined EMSS, NEP and 160 deg$^2$ 
high-redshift samples. The figure assumes $\Omega_{m0} = 1.0$, 
$\Omega_{\Lambda0} = 0.0$.}
\end{figure}

The first large sample comes from combining the 160 deg$^2$, EMSS and
NEP samples, denoted HEN. There are 109 objects with L(0.5,2.0) $\geq
1\times10^{43}$ $\mathrm erg\: s^{-1}$ and $0.3 \leq z \leq
0.85$. This size is comparable to the low-redshift samples.  There are
no clusters in common among the three samples and the overlap of their
survey regions on the sky is small, only $\sim5\%$, which has been
accounted for statistically. The HEN sample is large enough that a
simple comparison of the luminosity functions from it and REFLEX
demonstrates the evolution clearly, as Figure 5 shows.  A Schechter
function fits both the REFLEX and HEN samples well. Figure 6 gives the
1, 2 and 3 $\sigma$ contours for the slope and characteristic
luminosity, again demonstrating evolution at high statistical
confidence. At least the Schechter function normalization and slope
are evolving, but the charcteristic luminosity may not be. Comparison
of HEN with REFLEX shows a factor of 2 evolution at L(0.5,2.0) $=
2\times10^{44}$ $\mathrm{erg\:s^{-1}}$ over the redshift interval 0.11
to 0.45 ($\Omega_{m0} = 1.0$, $\Omega_{\Lambda0}= 0.0$).

\begin{figure}
\plottwo{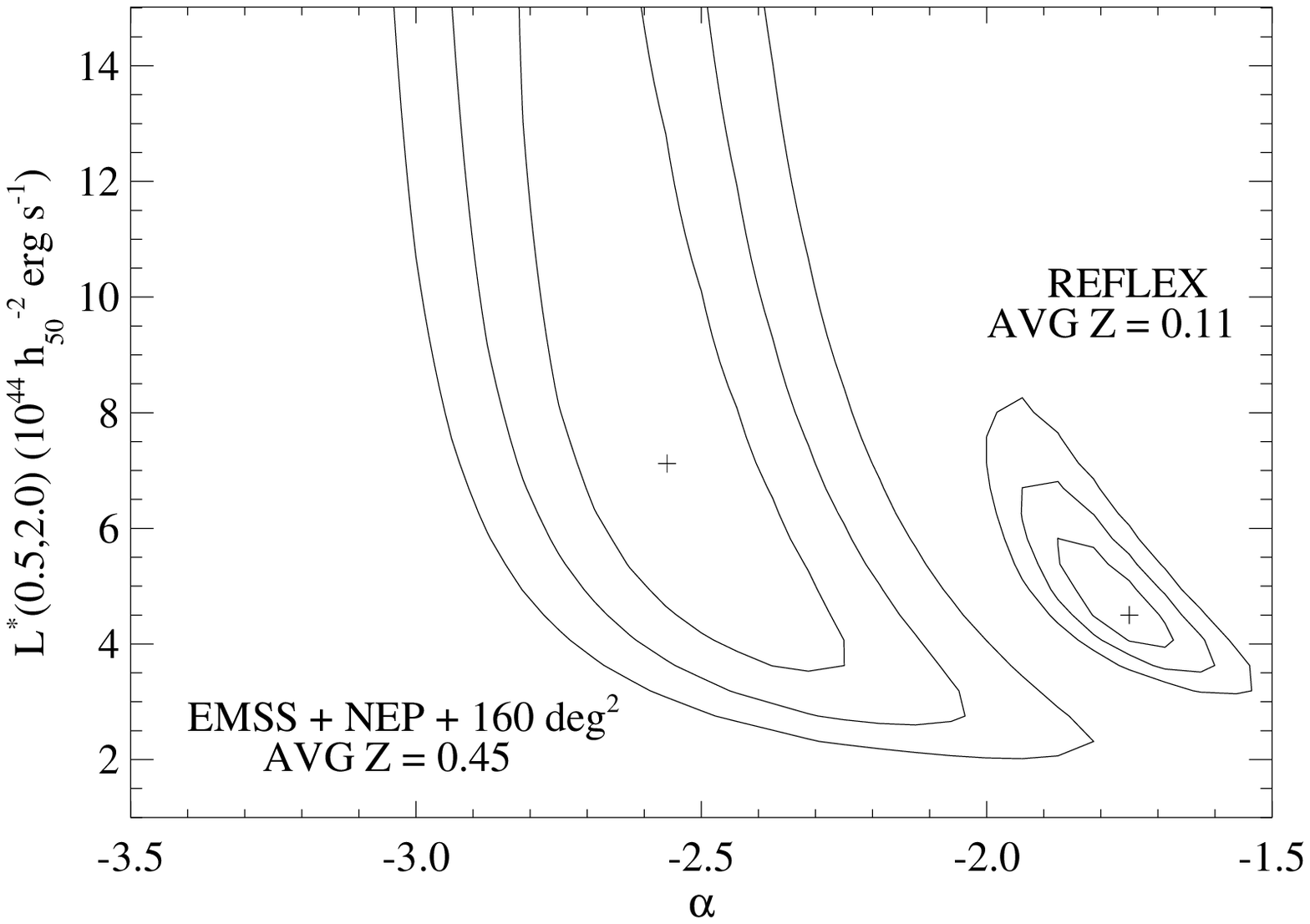}{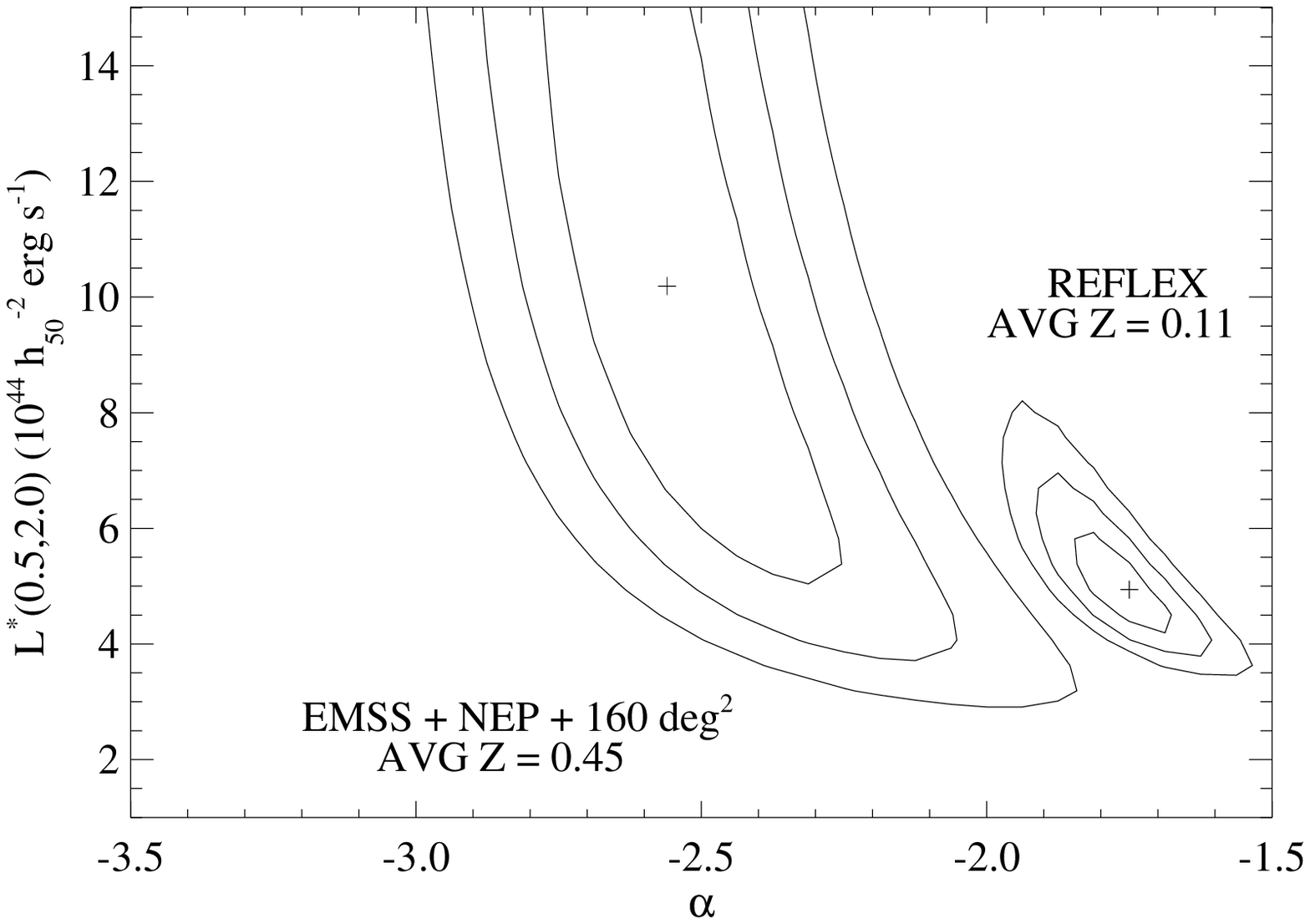}
\caption{The 1, 2 and 3$\sigma$ contours for the slope and characteristic
luminosity of individual Schechter function fits to the REFLEX and HEN
samples. a. (left) For a cosmology with $\Omega_{m0} = 1.0,
\Omega_{\Lambda0} = 0.0$. b. (right) For a cosmology with $\Omega_{m0}
= 0.3, \Omega_{\Lambda0} = 0.7$.}
\end{figure}

It is the nature of flux-limited samples that different luminosity
objects are predominately found at different redshifts. Thus the
luminosity functions in Figure 5 do not refer to unique epochs. This
situation may be overcome to some extent by fitting to an
evolving luminosity function, termed the AC model.  The AC model
generalizes the Schechter funtion to: $\mathrm {n(L,z) = n_0(z)
L^{-\alpha(z)} exp(-L/L^*)}$, with $\mathrm {n_0(z) = n_0 (1+z)^A}$
and $\mathrm{\alpha(z) = \alpha_0 (1+z)^C}$. A maximum likelihood fit
of the AC model to the unbinned luminosity and redshift pairs for the
REFLEX and HEN clusters gives the following: $\mathrm {n_0 =
4.65^{+0.58}_{-0.64} \times10^{-7} Mpc^{-3} (10^{44}
erg\:s^{-1})^{\alpha-1}}$, $\mathrm{L^* = 6.86^{+4.59}_{-1.96} \times
10^{44} erg\:s^{-1}}$, $\alpha_0 = -(1.70^{+0.15}_{-0.13})$, A =
-(1.17 $\pm$ 0.84), C = 0.90$^{+0.60}_{-0.35}$ ($\Omega_{m0} = 1.0$,
$\Omega_{\Lambda0}= 0.0$). The errors are at 68\% confidence for 4
parameters with the error on $\mathrm{n_0}$ determined from letting
the errors on the other 4 parameters take their extreme
values. Figure 7 gives the best fitting AC luminosity functions, now
at two particular epochs.

\begin{figure}
\plotfiddle{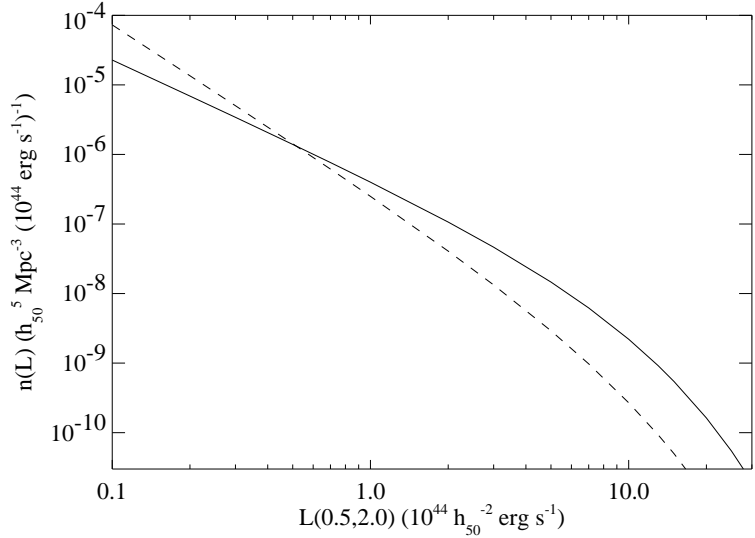}{2.5in}{0}{60}{60}{-190}{-230}
\caption{Best fitting AC model luminosity functions. The solid line
is z = 0.0, while the dashed line is z = 0.5. The figure assumes
$\Omega_{m0} = 1.0$, $\Omega_{\Lambda0} = 0.0$.}
\end{figure}

The second large high redshift sample is the Massive Cluster Survey or
MACS. This survey seeks a large sample of high luminosity clusters at
high redshift. About 300 objects are expected if there is no
evolution.  Presently the survey is $\sim 90\%$ complete and has 119
clusters.  Note that in Section 3.2 an additional $10\%$ systematic
error has been added in quadrqture to the MACS statistical errors.

\begin{figure}
\plotfiddle{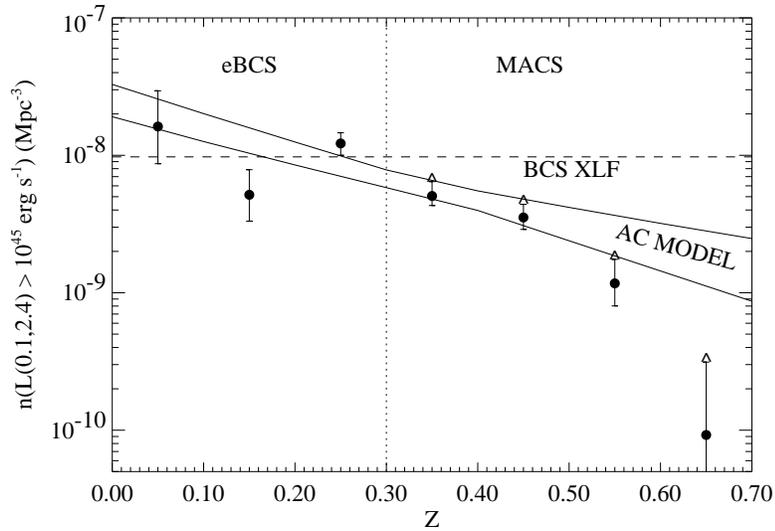}{2.5in}{0}{60}{60}{-190}{-230}
\caption{A nonparametric description of X-ray cluster luminosity 
evolution detected in the eBCS/MACS samples. The integral luminosity
function is shown as a function of redshift. Strictly speaking, the
MACS results are lower limits, but the estimated completeness of this
survey is $\sim90\%$. Also shown is the AC model with its $1\sigma$
error band. There are no adjustable parameters used to fit the AC
model to the eBCS/MACS. The figure assumes $\Omega_{m0} = 1.0$,
$\Omega_{\Lambda0} = 0.0$.}
\end{figure}

A final sample of $\sim130$ objects means that MACS detects evolution
also at high significance. Figure 8 shows a nonparametric
representation of this evolution, comparing the MACS and eBCS integral
luminosity functions. The comparisons in Section 3.2 folded a model
fit to the REFLEX luminosity function through each survey selection
function. MACS finds a factor of $\sim10$ evolution at L(0.1,2.4) $>
10^{45}$ $\mathrm{erg\:s^{-1}}$ over the redshift interval 0.05 to
0.55 ($\Omega_{m0} = 1.0$, $\Omega_{\Lambda0}= 0.0$).  Figure 8 also
shows that the AC model fits the MACS data well, with no adjustable
parameters and with few clusters in common between eBCS/MACS and
REFLEX/HEN.

Two high statistics, high redshift samples agree on the amount and
nature of cluster X-ray luminosity evolution. The previous apparent
disagreements were probably just poor statistics.

\subsection{Constraining Cosmology From Cluster Luminosity Evolution}

Given the discussion in the Introduction, the question naturally
arises whether cluster luminosity evolution may be used to determine
the cosmology of the universe.  The difficulty is the extra evolution
experienced by the gas in addition to that driven by gravity, as the
L - T relation shows. Since X-ray luminosity is proportional to the
square of the gas density, will this extra evolution be large enough
to mask that from gravity?  The weak evolution of the L - T relation
shows that the extra evolution occured before what is currently
observable, possibly allowing a simple correction. Comparing data to a
theory that incorporates the observed L - T (or L - M) relation into the
gravitational evolution predictions has in fact been used to constrain
cosmological parameters from cluster luminosities (eg Henry et al.,
1992; Borgani et al., 2001; Viana, Nichol \& Liddle, 2002).

\section{Cluster X-ray Temperature Evolution}

Perhaps a cleaner way to measure the evolution of cosmic structure
using X-ray observations of clusters is via cluster temperature
evolution. The temperature is more likely to reflect better the
underlying gravity driven evolution than the luminosity.

\subsection{Cluster X-ray Temperature Samples}

Clusters are not standard temperature temperature baths. Just as with
luminosity evolution, measuring temperature evolution requires
constructing temperature functions at two epochs. Again, statistically
complete samples are required at the two epochs. However it is
substantially more difficult to measure cluster temperatures than
luminosities. At least 1000 photons are required to provide a
reasonable temperature measurement. Obtaining large samples is
difficult. Obtaining even two independent samples at high redshift has
so far not been possible. Table 3 gives some properties of cluster
temperature samples. The last column of the table gives the numbers
of clusters with measured temperatures from ASCA and with temperatures
estimated from the L - T relation.

{\small
\begin{table}
\caption{Cluster Temperature Samples (z$<$0.3 All Sky; z$>$0.3 778 deg$^2$)}
\begin{tabular}{lclllc}
\tableline
Reference&Num&z Cut&T Cut&F(0.1,2.4)(cgs)&ASCA/L-T\\
\tableline
Edge et al. (1990)&46&&&$>1.7\times10^{-11}$&0/2\\
Henry \& Arnaud (1991)&25&$<0.17$&&$>3.0\times10^{-11}$&25/0\\
Markevitch (1998)&30&0.04-0.09&&$>2.0\times10^{-11}$&26/4\\
Blanchard et al. (2000)&50&$<0.10$&&$>2.2\times10^{-11}$&34/0\\
Pierpaoli et al. (2001)&38&0.03-0.09&$>3.5$&L-T, z&?/1\\
Ikebe et al.  (2002)&61&&$>1.4$&$>2.0\times10^{-11}$&56/2\\
Henry (2002) EMSS&22&0.30-0.85&&$>1.2\times10^{-13}$&21/1\\
\tableline
\tableline
\end{tabular}
\end{table}
}

The low redshift samples are comprised of bright well-known clusters
and are all more or less the same. The Edge et al. and Henry \& Arnaud
samples come from non-imaging surveys, the rest of the low redshift
samples come from ROSAT. The Markevitch and Ikebe et al. samples have
corrections for cooling centers; about 20\% of the Blanchard et al.
sample has these corrections. The high redshift clusters are
unresolved by ASCA, which, when combined with the low statistics, does
not permit corrections for cooling centers. Figure 9 shows the
integral temperature functions at two epochs. The same factor $\sim2$
evolution seen in the luminosities is also seen in the temperatures.

\begin{figure}
\plotfiddle{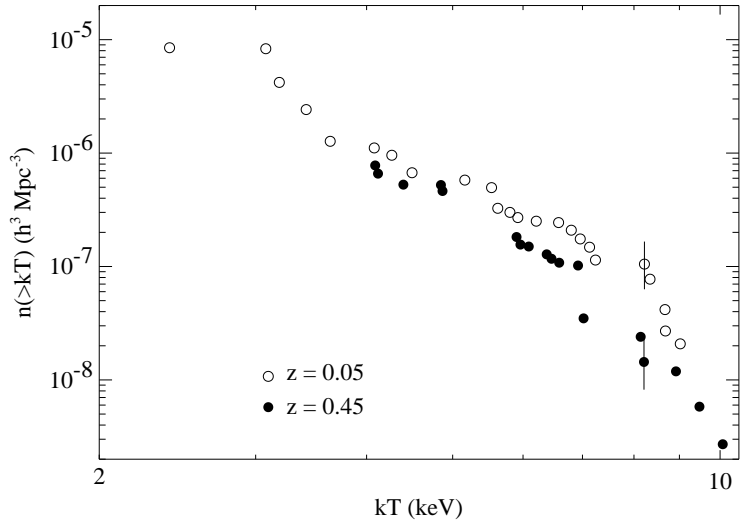}{2.5in}{0}{60}{60}{-190}{-230}
\caption{A nonparameteric, unbinned presentation of the cluster integral
temperature function from the Henry \& Arnaud (1991) and EMSS samples
(Henry, 2002). The figure assumes $\Omega_{m0} = 1.0$,
$\Omega_{\Lambda0} = 0.0$.}
\end{figure}

\subsection{Cosmological Results From Cluster X-ray Temperature Evolution}

Deriving cosmological constraints from X-ray observations is a
two-step process. The fundamental variable of the theory is an
object's mass so the process starts with a mass function. The analytic
mass function that best matches the large N-body simulations comes
from Sheth and Torman (1999). This function agrees better with the
simulations than does the venerable Press - Schechter (1974) function.
However, the differences are only a factor of $\sim2$. 

The evolution of the mass function is very sensitive to cosmological
parameters, but the mass of a cluster is difficult to determine. So,
the mass must be converted to a more easily observed quantity, in this
case the temperature. A top hat collapse is most often used to do
that. This theory gives $\mathrm{kT=1.42\beta^{-1}_{TM}[\Omega_{m0}
\Delta(\Omega_{m0},\Omega_{\Lambda0},z)]^{1/3}
(hM_{15})^{2/3}(1+z)}$. Here $\mathrm{\beta_{TM}}$ is a factor near
unity that accounts for the effects of partial virialization of the
cluster. Its exact value comes from hydrodynamical
simulations. $\mathrm{\Delta(\Omega_{m0},\Omega_{\Lambda0},z)}$ is the
ratio of the average cluster mass density to the background mass
density for a cluster that collapses at redshift z. It has recently
been determined for arbitrary $\Omega_{m0}$ and $\Omega_{\Lambda0}$ by
Pierpaoli et al. (2001). M$_{15}$ is the cluster mass in units of
$10^{15}M_{\sun}$.

One final issue concerns the selection function. X-ray clusters are
selected by flux, not temperature. That is the solid angle surveyed to
a given flux limit is known. Pierpaoli et al. (2001) circumvent this
by using the L - T relation to insure that their temperature and
redshift selection was compatible with the flux limits of existing
samples. Usually, the flux limit is recast as a limit in the
luminosity-redshift plane and then the luminosity is converted to a
temperature via the L - T relation. The dispersion in the L - T
relation is included by averaging over it which gives the solid angle
surveyed for a given cluster of temperature kT at redshift z:
\[\mathrm{\Omega(kT,z) =
\int\frac{dlog(L)\:\Omega(L,z)}{\sqrt{2\pi\sigma^2(logL)}}
exp-(log(C[kT]^\alpha)-logL)^2/2\sigma^2(logL)}\]

There has been a great deal of recent work in this area. Among the many
results are: Bahcall \& Fan (1998), Donahue \& Voit (1999), Henry (2000;
2002), Pierpaoli et al. (2001), and Viana \& Liddle (1999).

Constraints on cosmological parameters presented here come from a
maximum liklihood fit of the evolution theory to the unbinned
temperature and redshift data for the samples in Figure 9. There are
four interesting parameters in the theory: the matter and dark energy
densities and the shape and normalization of the matter power
spectrum. The effects of temperature errors are included in the
likelihood function. Figures 10 and 11 show some of the results
(Henry, 2002).

\begin{figure}
\plotfiddle{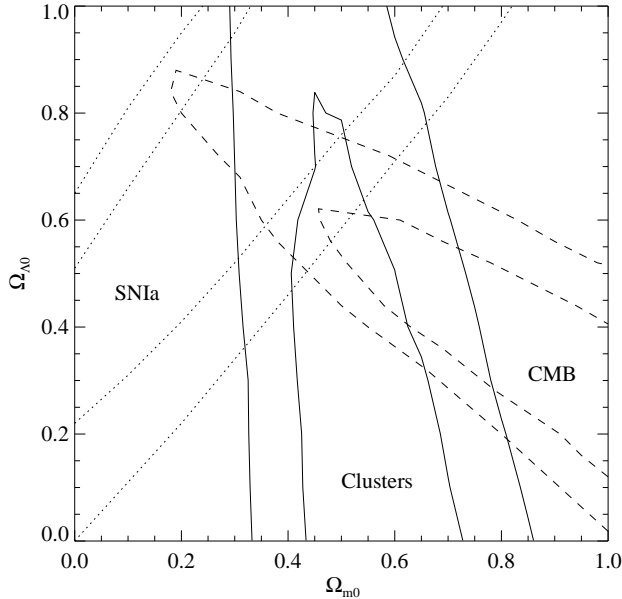}{2.8in}{0}{55}{55}{-170}{-170}
\caption{The 1 and 2 $\sigma$ contours  for $\Omega_{m0}$ and 
$\Omega_{\Lambda0}$ from cluster temperature
evolution, cosmic microwave background and supernovae. The supernovae
and CMB constraints are from Perlmutter et al. (1999a) and Efstathiou
et al.  (2002) respectively.}
\end{figure}

\begin{figure}
\plottwo{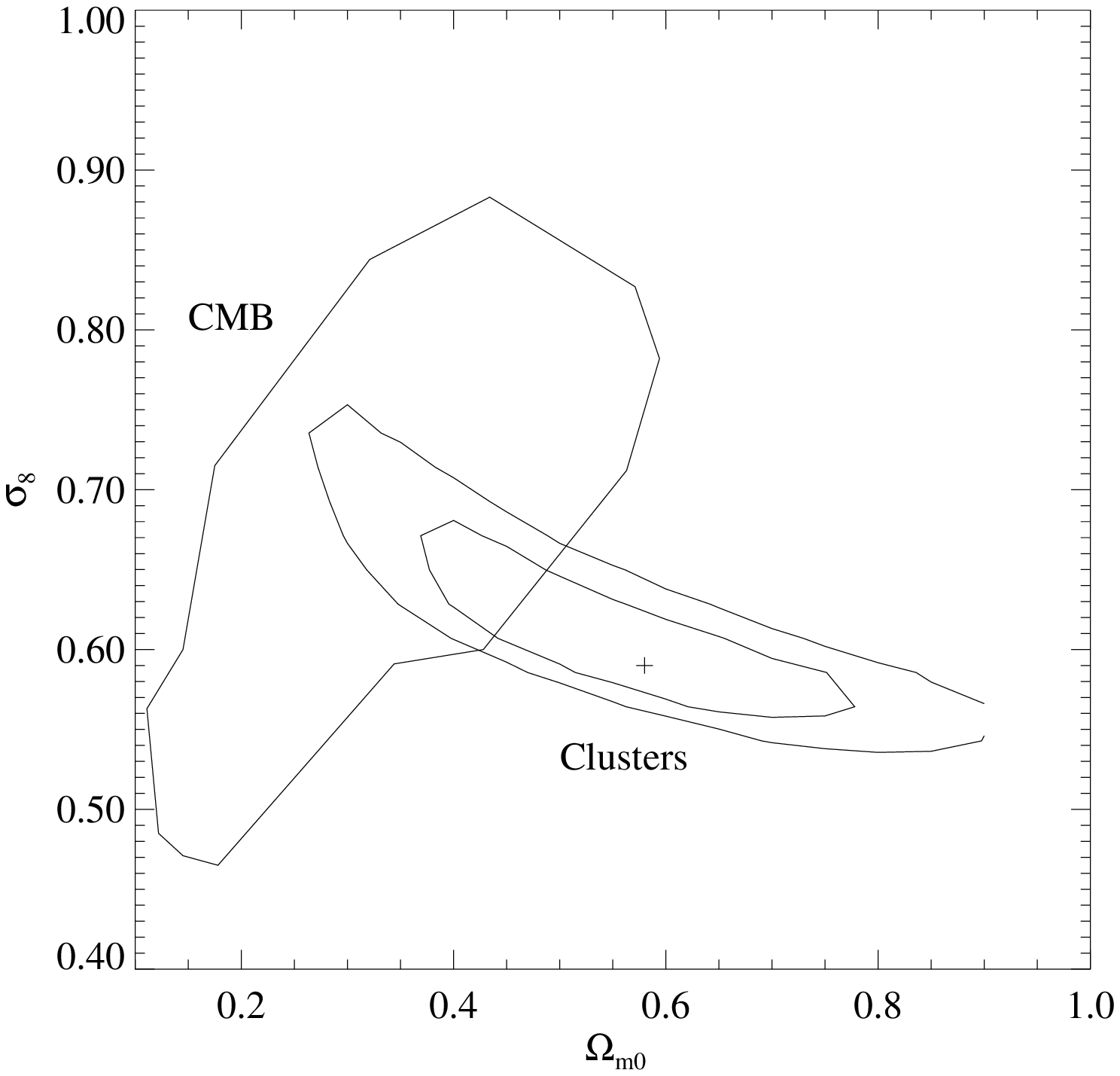}{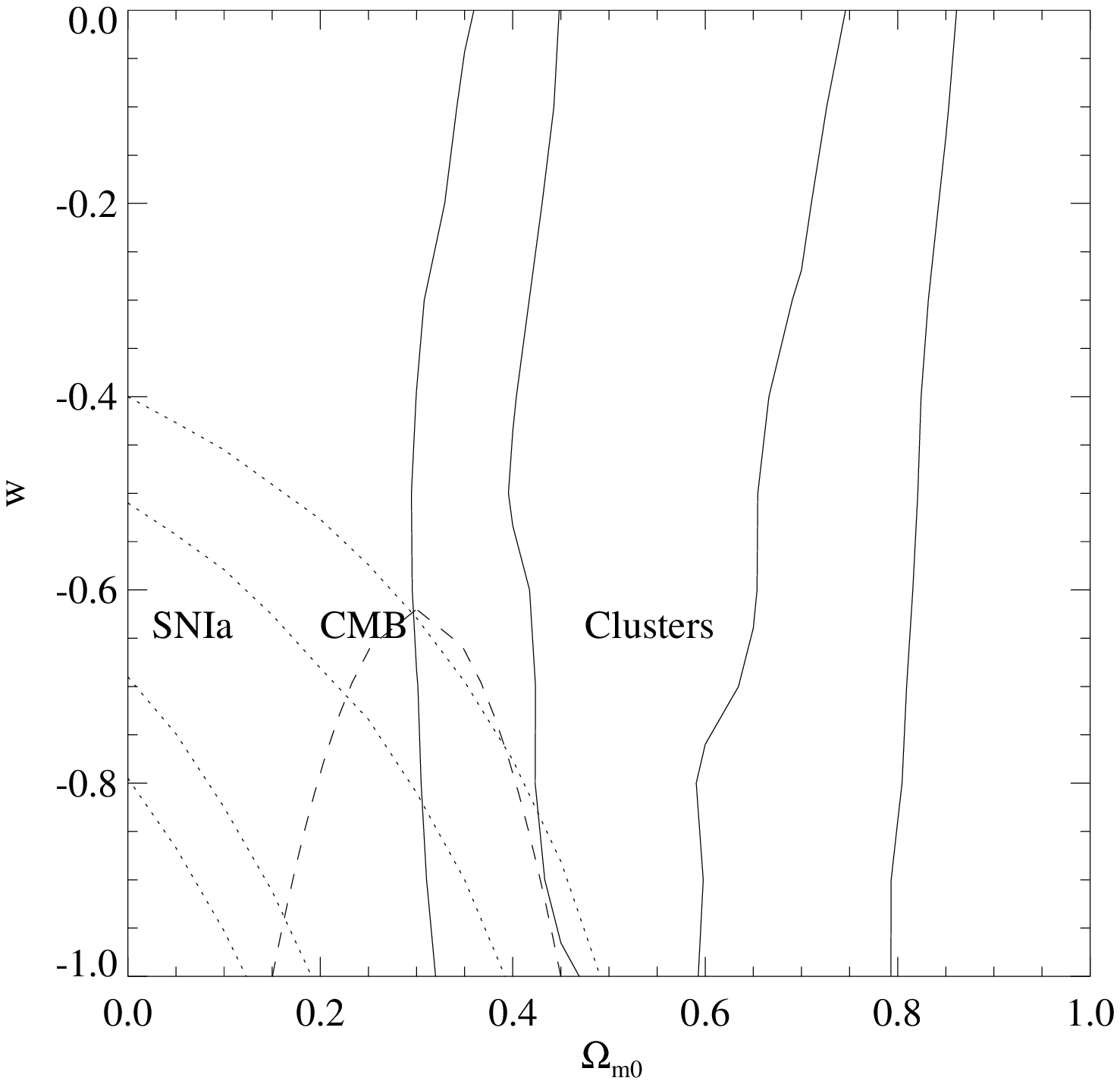}
\caption{a. (left) Constraints on $\mathrm{\sigma_8\:and\:\Omega_{m0}}$ 
shown as 1 and 2 $\sigma$ contours from cluster evolution and a 2 
$\sigma$ contour from the CMB (Melchiorri \& Silk, 2002). The CMB 
constraint assumes h = 0.71 $\pm$ 0.07 and $\Omega_{m0}$ + 
$\Omega_{\Lambda0}$ = 1. b. (right) 
Constraints on w and $\Omega_{m0}$ shown as 1 and 2 $\sigma$ contours 
from cluster evolution and supernovae and a 1 $\sigma$ contour from 
the CMB. The supernovae and CMB constraints are from Perlmutter, 
Turner \& White (1999b) and Bean \& Melchiorri (2001), respectively.
All results assume that $\Omega_{m0} + \Omega_{Q0}$ = 1 and the
CMB result further assumes h = 0.72 $\pm$ 0.08.}
\end{figure}

Figure 10 shows that clusters provide a band of constraints in the
$\Omega_{m0}\:-\:\Omega_{\Lambda0}$ plane of comparable width but
different orientation to those from supernovae and microwave background
observations. Further, all three bands intersect at the same point in
the plane. $\Omega_{m0} = 0.4$, $\Omega_{\Lambda0} = 0.6$ is
consistent with all three data sets at the $\sim 1 \sigma$ level. For
example, Rubino-Martin et al. (2002) show that all CMB data reported
to date, plus the supernovae data from Perlmutter et al. (1999a), plus
assuming h = 0.4 - 0.9 yields $\Omega_{m0}=0.35\pm0.04$ and
$\Omega_{\Lambda0} = 0.68\pm0.04$ at 68\% confidence.

The normalization of the power spectrum, $\sigma_8$, agrees with the
microwave background normalization evolved to the present, Figure
11a. Thus the fluctuations seen in the CMB really do grow to form the
clusters seen today.  The present cluster sample size is not large
enough to provide tight constraints on w, Figure 11b. However, w = -1
is a good bet considering the constraints provided by supernovae and
the CMB.

\subsection{Cluster X-ray Temperature Evolution Systematics}

It is of course important to consider the effects of systematic errors
on the constraints derived in the last section. The three effects
considered here are: 1. Using the older Press - Schechter (1974) mass
function; 2. Using an empirical normalization of the mass -
temperature relation, i.e. $\mathrm{\beta_{TM}}$, from Finoguenov,
Reiprich \& B\"ohringer (2001) instead of a theoretical normalization;
3. Setting all temperature errors to 2\% or the actual value,
whichever is less.  All systematics effects considered change the
constraint on $\sigma_8$ by $\sim1\sigma$, the constraint on
$\Omega_{m0}$ by $\sim0.5 \sigma$ and the constraints on
$\Omega_{\Lambda0}$ and w by $\ll 1\sigma$. Thus statistical errors
appear to dominate the error budget with the present sample size of
about fifty objects.

\section{Conclusions}

The L - T relation shows that the hot X-ray cluster gas experienced a
more complicated thermodynamic history than that resulting from the
cluster collapse. There was possibly some heating prior to collapse
and/or cooling with lesser heating after collapse. 

There is no longer a question whether cluster X-ray luminosities
evolve, they do. Seven nearly independent surveys show some evolution
and systematics do not appear to be a big effect. Large high redshift
samples, i.e. those containing more than 100 clusters, are needed to
characterize the evolution. Two such samples currently exist. HEN
shows a factor of 2 evolution compared to REFLEX at L(0.5,2.0) $=
2\times10^{44}$ $\mathrm{erg\:s^{-1}}$ over the redshift interval 0.11
to 0.45 ($\Omega_{m0} = 1.0$, $\Omega_{\Lambda0}= 0.0$). MACS shows a
factor of 10 evolution compared to eBCS at L(0.1,2.4) $> 10^{45}$
$\mathrm{erg\:s^{-1}}$ over the redshift interval 0.05 to 0.55
($\Omega_{m0} = 1.0$, $\Omega_{\Lambda0}= 0.0$).

A cosmology with $\Omega_{m0} = 0.4$, $\Omega_{\Lambda0}= 0.6$ is
consistent with cluster X-ray temperature evolution, the cosmic
microwave background and supernovae at the $\sim1\sigma$ level. The
systematics of cluster temperature evolution constraints on cosmology
are a minor concern at present, i.e. they are about the same size as
the statistical errors.  However, this method is so statistically
powerful that systematics will be the dominate effect as soon as the
sample sizes become only a factor of two or three larger, at least
for the determination of $\sigma_8$. 

A more positive statement would be that an X-ray cluster survey of
10,000 deg$^2$ to a flux of $\mathrm{F(0.5,2.0) > 5\times10^{-14}
\:erg\:cm^{-2}\:s^{-1}}$ would yield $\sim18,000$ clusters to z
$\sim1.5$ and provide constraints of similar statistical quality as
the upcoming Planck and the proposed SNAP missions (Petre et al.,
2001). In fact the great complementarity of cluster evolution, cosmic
microwave background, and supernovae constraints exhibited in Figure
10 would provide a check on the systematics of all three methods.

\acknowledgements Thanks are due to the organizers of our conference for
one of the best meetings I have ever attended. Particular thanks go to
the organizers of the special night sessions that were also great.  I
want to acknowledge my many collaborators with whom I have worked on
cluster surveys over the years. These include: I. Gioia, C. Mullis,
W. Voges, U. Briel, H. B\"ohringer and J. Huchra for the NEP;
A. Vikhlinin, C. Mullis, I. Gioia, B. McNamara, A. Hornstrup,
H. Quintana, K. Whitman, W. Forman, and C. Jones for the 160
deg$^{2}$. Most of the work described here will eventually appear as
publications coauthored with them. I thank H. Ebeling and L. Jones who
provided results prior to publications for MACS and WARPS II and
H. B\"ohringer who provided the digital data from Figure 20 of
B\"ohringer et al. (2001).

\end{document}